\documentstyle[prl,aps,multicol,epsf]{revtex}
\begin{document}

\draft
\title{
  Does the Heisenberg model describe  the multimagnon spin dynamics 
   in antiferromagnetic CuO layers ?  }

\author{J. Lorenzana$^{1,2}$, J. Eroles$^{2,3}$ and S. Sorella$^4$}
\address{
$^1$Istituto Nazionale di Fisica della Materia - Dipartimento 
di Fisica, Universit\`a di Roma ``La Sapienza'',
 Piazzale 
A. Moro 2, I-00185 Roma, Italy.\\
$^2$Centro At\'{o}mico Bariloche and Instituto Balseiro, 
8400 S. C. de Bariloche, Argentina.\\
$^3$ Theoretical Division, Los Alamos National Laboratory, New Mexico 87545  \\
$^4$ Istituto Nazionale di Fisica della Materia, SISSA, 
Via Beirut 4, 34014 Trieste, Italy  }
\date{\today}
\maketitle
\begin{abstract} 
 We compute the absorption spectrum for  multimagnon excitations assisted
 by phonons in insulating layered cuprates using exact diagonalization in 
clusters of up to 32 sites. The resulting line shape is very sensitive 
to the underlying magnetic Hamiltonian describing the spin dynamics.
For the usual Heisenberg description of undoped Cu-O planes we find,
in  accordance with experiment,  a two-magnon peak followed by high energy 
side bands. However the relative weight  of the side bands  is
too small to reproduce the experiment. An extended Heisenberg model
including a sizable four-site cyclic exchange term is shown to be 
consistent  with the experimental data.

\end{abstract}
\begin{multicols}{2}

\narrowtext

The starting point of many theories describing high-temperature 
superconducting cuprates is the undoped parent compound. A consistent 
description of
this phase is of great importance since the usual approach 
 is to ``extend'' this model to describe the doped phase like the t-J 
model\cite{zha88} or the spin fermion  model\cite{bar95}. 
 Also the  understanding of this phase, namely the 
physical realization of a  two-dimensional (2d)
 spin-1/2 quantum antiferromagnet,  is a fundamental problem in itself. 

It is usually assumed that the 2d Heisenberg model (HM) 
\begin{equation}
  \label{eq:hei}
  H_{\rm Hei}\equiv \sum_{i,j}J_{ij} \bbox{S}_i.\bbox{S}_j,
\end{equation}
describes the physics of the stoichiometric materials\cite{man91} with 
$J_{ij}=J$ for the first nearest neighbors and zero otherwise. 

In this work we  show that infrared (IR) optical
absorption spectra due to phonon assisted multimagnon excitations 
 is  very sensitive to the magnetic Hamiltonian.  We find
that the usual Heisenberg description is incompatible
with IR experimental data when more than two magnons are involved.
 An extended Heisenberg model 
including further neighbor interactions in $J_{ij}$ and 
 a   four-spin  cyclic-exchange (4SCE)  
term\cite{gag90a,tak77,rog89,sch90,hon91,hon93,lem97ero99} is shown to 
explain the IR experiments. The same model has been shown 
to be consistent  with other spectroscopic data\cite{lem97ero99}.

The  4SCE was introduced by Takahashi\cite{tak77}  and  by 
 Roger and Delrieu on the present context,  using a
 4th order perturbative analysis\cite{rog89}. It was also supported by exact 
 diagonalization (ED) studies of a multiband-Hubbard model\cite{sch90} 
describing the Cu-O planes. The main effect of this term in the Hamiltonian is 
to permute cyclically four spins on a plaquette.

Though  small exchange interactions,  going beyond the first neighbors 
are expected\cite{sin89},  the 4SCE term is not generally 
accepted in the literature\cite{man91}. 
 Up to now there has not been any clear experimental way to 
rule out or  confirm  the presence of this term. 
In fact the experimentally established\cite{man91} 
 single-magnon spin-wave branch
is rather insensitive, at low energies, to the presence of the 4SCE 
term\cite{sch90,chu92}.  

 Any noticeable effect of the 4SCE term is expected to occur at high 
energies which are  accessible in  optical experiments. One example 
 is magnetic Raman (MR) light-scattering\cite{ell69}. 
 The line shape has an asymmetric peak close to
$3J$ due to two-magnon excitations and a shoulder at higher energy which 
is believed to arise  from a  four-magnon process\cite{sug90}.
 The width of the 
two-magnon peak and the 
four magnon shoulder are anomalous in the sense that they do not agree with 
a conventional interacting spin-wave theory description of the line 
shape\cite{rog89,sch90,hon91,hon93,lem97ero99,sin89,kno90,nor95}.
Theoretical studies have attributed both anomalies to the presence 
of other terms in the 
Hamiltonian\cite{rog89,sch90,hon91,hon93,lem97ero99,kno90,nor95} including
 the
4SCE term\cite{rog89,sch90,hon91,hon93,lem97ero99}. Although the latter 
assignment is encouraging we argue that the
analysis of the MR line shape is not conclusive 
(see Ref.~\cite{note}).

\begin{figure}[tbp]
\epsfverbosetrue
\epsfxsize=7cm
$$
\epsfbox{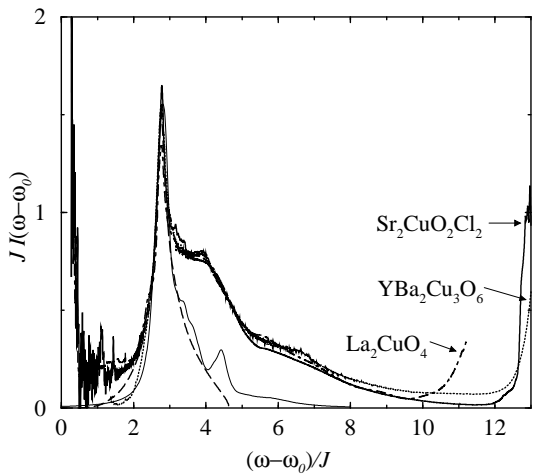}
$$
\caption{  $I(\omega-\omega_0)$ 
obtained from the experimental  absorption using  
Eq.~(\protect\ref{adwe}). Following Refs.~\protect\cite{per93,per98} a 
linear background has been subtracted form the raw absorption data. 
We show the data for
 La$_2$CuO$_4$\protect\cite{per93}, Sr$_2$CuO$_2$Cl$_2$\protect\cite{per93} 
and YBa$_2$Cu$_3$O$_6$\protect\cite{gru96} in a dimensionless  scale
using the following  reference energies  
 $(\omega_0,J)=$ (0.08,0.12), (0.06,0.11) and (0.06,0.1)
eV respectively.  The energy scale $J$ 
and the intensity scale were adjusted to match the first
peak in ED. 
We also show the HM theoretical line shape in ED  (thin line) 
for a 32 site cluster and a Lorentzian broadening of 0.16 $J$ 
and in interacting spin-wave  theory\protect\cite{lor} 
(dashed line). The intensity scale of the latter was adjusted to match the
ED intensity. 
}
\label{idwf}
\end{figure}

Another experiment, which probes the multimagnon response, is
phonon-assisted multimagnon light absorption  
(PAMLA)\cite{per93,per95,per98,gru96,lor,suz96}. 
In this experiment an absorbed photon simultaneously creates a phonon
and a multimagnon excitation. The absorption mechanism is well
understood\cite{lor,suz96} allowing to
make theoretical predictions on the  nickelates\cite{lor}(b) 
(2d, spin 1) which  were  successfully corroborated\cite{per95,per98}.  
Besides, the line shape was 
computed for spin-1/2 1d Cu-O systems\cite{suz96}  with great 
accuracy\cite{lor}(d). The experimental line shape is reproducible even among
different materials (see Fig.\ref{idwf}).   All this puts the interpretation 
of the
IR line shape on a firmer basis than the MR line shape\cite{note} 
making it an ideal candidate to test models of the spin dynamics. 

For the 2d cuprates the line shape has been measured in several
materials\cite{per93,per95,per98,gru96}. In 
Fig.~\ref{idwf} we show the line shape for three different cuprate compounds
in a dimensionless scale.  All the   data collapse to a unique
curve, implying that this experiment  depends only on the common CuO 2d layers.
The high energy upturn is due to the charge transfer band and the low
energy upturn is due to the phonons.

The line shape has, like in MR scattering,   
a structure close to $3J$ (measured from the  phonon energy) due to the 
two-magnon process. In addition strong side-bands appear at
high energies.

A recent study has  suggested that, though  the main peak is of magnetic 
origin, the side bands may be explained by the presence of a
  d-d exciton\cite{per98}. In this case, since a d-d exciton will 
depend on details outside the Cu-O planes  one would expect the position
and intensity of the side bands to be unrelated to the position and 
intensity of the main peak when different materials are examined.
 Instead the scaling 
shown in Fig.~\ref{idwf} shows that this is not the case ruling out the
 exciton explanation. We anticipate that the
main peak and the side bands can be understood  within the PAMLA
mechanism if the appropriate magnetic Hamiltonian is used.

The PAMLA absorption coefficient  is given by,
\begin{equation} 
\alpha(\omega)  =\alpha_0 \omega I(\omega-\omega _{0})
\label{adwe}
\end{equation}
where $\alpha_0$ is a material dependent constant defined in 
Ref.~\cite{lor}, $\omega _{0}$ is the frequency of the stretching mode 
phonon. $I(\omega)$ is the weighted sum of a two-spin spectral
function  on the Brillouin zone\cite{lor} 
\begin{eqnarray}
I(\omega)&=&\frac{8}{N}
\sum_{\bbox{q} } \sin(q_x/2)^2( \sin(q_x/2)^2+\sin(q_y/2)^2) \times\nonumber\\
& &\sum_{\nu }
\left| \langle 0 |B_{\bbox{q}}^x |\nu \rangle \right|^{2}
\delta(\omega+E_{0 }-E_{\nu })
\end{eqnarray}
and we introduced the Fourier transform 
$  B_{\bbox{q} }^x=\frac1{\sqrt{N}}\sum_j e^{i \bbox{q . R}_j} 
\bbox{S_{R_j}}.\bbox{S_{R_j+\hat{x}}}. $

The two-magnon peak has been accurately fitted with a two-magnon
interacting spin-wave theory computation\cite{lor}(a),(b) in the
HM which we report  on Fig.~\ref{idwf}. 

The physics of the two-magnon peak and the side bands can be understood 
with a simple argument\cite{hon93,lor}. If one approximates the
ground state by the classical N\'eel state the effect of the $B_{\bbox{q}}^x$
operator is to flip two spins  in nearest neighbor sites.
 The energy of this
excitation is $3J$ which is close to the energy of the peak observed. 
In the HM this state is
not an eigenstate and will mix with states with four, six,  etc. spin
flips. Consequently the spectral function  will show side bands at 
the energy of these excitations which in the Ising limit are
$4J$, $5J$, etc.

Since the side bands involve more than two magnons they cannot be
described  in the {\em two-magnon}-interacting spin-wave theory 
computation of Ref.~\cite{lor}(a),(b). 
It is therefore important to avoid  the spin-wave 
approximation  and establish with an unbiased technique 
 whether or not the side bands can be described by the 
HM. To  this purpose we have computed the
spectrum using ED on finite clusters.

\begin{figure}[tbp]
\epsfverbosetrue
\epsfxsize=7cm
$$
\epsfbox{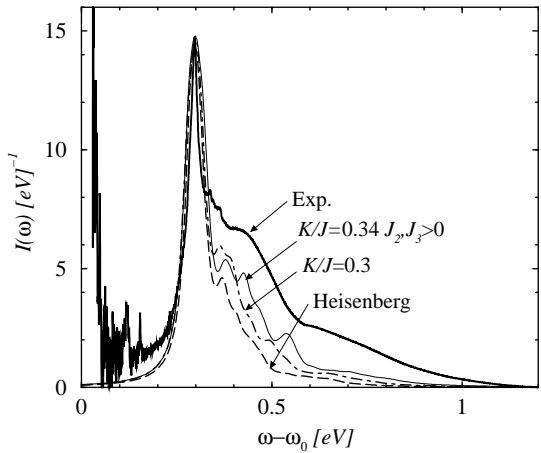}
$$
\caption{$I(\omega)$ in ED for a 26 site cluster and with
 the same parameter 
sets as in Fig.~\ref{momdn}. We also show the experimental line shape in
 Sr$_2$CuO$_2$Cl$_2$\protect\cite{per93} (thick line).
 The intensities were scaled to coincide at  the maximum 
 of  the  HM spectrum. In order to have 
the maximum  of all the spectra at the same $\omega$ 
we have used   different values of $J$ for the three different models:
 $J=0.1$eV (HM) $J= 0.12$ eV (4SCE with K/J=0.3)  
$J=0.15$ eV (EHM)  }
\label{idw2}
\end{figure}

 In Figs.~\ref{idwf},\ref{idw2} we show the  exact spectrum in 
different size clusters.  
Although the Heisenberg line shape has  some structure at the energy
of the side bands the relative intensity is much smaller than the
experimental one. Notice the similarity with the two-magnon spin-wave 
theory line shape (dashed line in Fig.~\ref{idwf}).

To make a more quantitative comparison  we have computed the 
cumulants\cite{sin89},
\begin{equation}
  \label{eq:m}
  (M_n)^n=\int (\omega - \rho)^n I(\omega) d\omega/I_T 
\end{equation}
with $I_T=\int I(\omega) d\omega$ and $\rho=\int \omega I(\omega) d\omega/I_T$.

$M_2$ and $M_3$ measure the width and asymmetry of $I(\omega)$ 
respectively ($M_1=0$).
 They are are obviously very sensitive to the presence of the side bands so it 
is natural to use them to characterize the line shape.

\begin{figure}[tbp]
\epsfverbosetrue
\epsfxsize=7cm
$$
\epsfbox{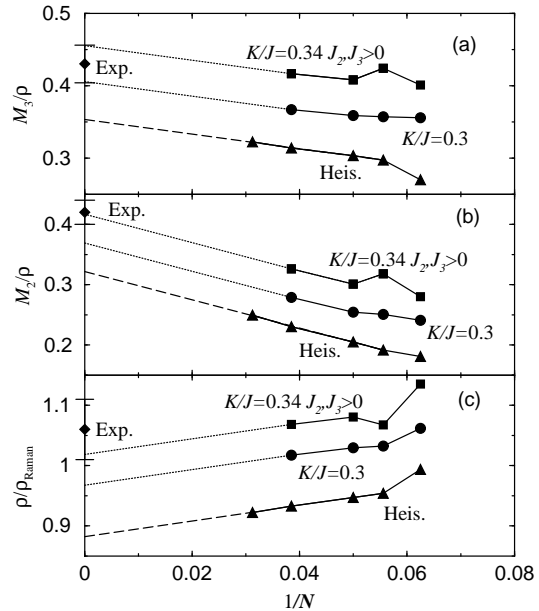}
$$
\caption{a) and b) $M_i/\rho$ as a  function of the inverse system size.
 c) $\rho/\rho_{\rm Raman}$\protect\cite{elsewhere}.  We show ED data 
for the HM (triangles), for the HM with  4SCE and $K/J=0.3$ (circles) 
and in an EHM with $K/J=0.34$, $J_2/J=0.04$,  $J_3/J=0.03$
(squares). The dashed line is a linear $1/N$  extrapolation for $N>16$ 
in the HM. The dotted lines are guides to the eye traced
parallel to the Heisenberg extrapolation. 
 The experimental values are computed from the  Sr$_2$CuO$_2$Cl$_2$
 data\protect\cite{per93,blu96}. 
The error bar reflects indeterminacies in background subtraction and, for 
$\rho_{\rm Raman}$, the dependence on excitation
 energy\protect\cite{blu96}.}    
\label{momdn}
\end{figure}

In Fig.~\ref{momdn} we show $M_2$ and $M_3$ 
and the ratio of the average energy $\rho$ to
the same quantity in MR ($\rho_{\rm Raman}$) 
for different  system size and for the experiment.

An extrapolation to an infinite system  confirms what 
 Figs.~\ref{idwf},\ref{idw2} suggest: the HM alone can not 
correctly describe this experiment. It is difficult to ascribe this 
to a failure of the PAMLA mechanism itself since, as we mention above,  
the mechanism has been 
successfully tested in an isostructural system with spin one, namely
La$_2$NiO$_4$\cite{lor}(b)\cite{per95}  and a system with the same spin (1/2)
but lower dimensionality\cite{lor}(d)\cite{suz96}. We therefore analyze 
a more realistic Hamiltonian to describe the spin dynamics.

The Hamiltonian with the 4SCE term reads $H=H_{\rm Hei}+H_{\rm 4S}$ with
\begin{eqnarray}
  \label{eq:4s}
    H_{\rm 4S}\equiv&K&\sum_{<i,j,k,l>}
 (\bbox{S_i}.\bbox{S_j})(\bbox{S_k}.\bbox{S_l}) +
 (\bbox{S_i}.\bbox{S_l})(\bbox{S_j}.\bbox{S_k})\nonumber\\&-&
 (\bbox{S_i}.\bbox{S_k})(\bbox{S_j}.\bbox{S_l})
\end{eqnarray}
where $<i,j,k,l>$ stands for the sum over groups of four spins on a plaquette.
This term can be shown to produce the cyclic permutation of the four
spins on the plaquette plus ordinary two spins exchanges of all the pair of 
spins of the plaquette including the ones on  the diagonals 
(see Ref.~\cite{hon93,rog83}). 
The parameter $K/J$ has been estimated using an ED 
mapping from a multiband Hubbard model\cite{sch90} to be around $0.3$.

In Fig.~\ref{idw2} we show the exact line shape in the HM and in the  model 
with the 4SCE term in a 26 site cluster.

We notice a strong sensitivity of the spectra to the 4SCE term. 
 The main effect is to transfer weight from the first
peak to the side bands. We already see in this limited size
cluster that the agreement with the experimental data is improved.
As a consequence of this transfer of spectral weight one sees an increase
of moments (Fig.~\ref{momdn}).

A fourth order perturbative analysis  of the Hubbard model\cite{lem97ero99}
 generates further nearest neighbor exchange interactions, beyond the 4SCE term
already considered. The additional terms are
spin exchange interactions for next-nearest neighbor sites ($J_{ij}=J_2$)
and for next-next-nearest neighbor sites ($J_{ij}=J_3$) in Eq.~(\ref{eq:hei}).
In order to be systematic we consider therefore an extended Heisenberg model 
with all magnetic interactions arising at fourth order.

To estimate the parameters we followed Ref.~\cite{lem97ero99} and apply 
the perturbative expressions to an extended Hubbard model including 
 second ($t_{2nd}/t_{1st}=0.15$)  and third   ($t_{3rd}/t_{1st}=-0.12$) 
neighbor hoppings $t$. The Hubbard $U$ is taken as $8 t_{1st}$.  
This approach has been shown to be  consistent with  
 MR, neutron-derived spin wave velocity as well as 
angular-resolved-photoemission spectroscopy data\cite{lem97ero99}.
From now on we refer to the resulting magnetic Hamiltonian as extended
 Heisenberg model (EHM).

In Fig.~\ref{idw2} we show the line shape and in Fig.~\ref{momdn} we show
the size dependence of the moments. 
We see that the effect of these terms also improves the agreement. Again we 
notice a strong sensitivity of the spectra to the underlying magnetic 
Hamiltonian.

A fine tuning of the Hamiltonian parameters,  which   accurately 
reproduce the experiments,  is not possible  with the 
limited sizes available, due to the difficulty  of 
 a precise extrapolation of the cumulants to an infinite system. 
Clearly a sizable value of $K/J$ is needed and probably also a non negligible 
$J_2/J$ and $J_3/J$. Our best estimate is the latter parameter set considered. 
In fact a rough extrapolation from Fig.~\ref{momdn}, assuming a similar 
scaling as for the HM, gives a result quite close to the experimental data. 
We mention that by setting $K=0$ in the EHM we where not
 able to obtain an acceptable fit to the experimental data with reasonable 
values of $J_2/J$ and $J_3/J$.
 Even for the EHM Fig.~\ref{idw2} does not show 
perfect agreement  between theory and experiment but the moment analysis 
shows that this can be ascribed  to a finite size effect.

The energy scale can be fixed by matching  the first moment 
 with the experimental first moment. In the EHM we find 
$J\sim 0.19 eV$  for Sr$_2$CuO$_2$Cl$_2$ in good agreement 
with the value found in  Ref.~\cite{lem97ero99} using other spectroscopic 
data. As an alternative procedure one
can adjust the position of the first peak as done in  Fig.~\ref{idw2}.
This gives a somewhat  smaller value probably due to finite size effects. 
   These values of $J$ are not in contradiction with the smaller value of 
$J$ usually quoted in cuprates
($J \sim 0.1 \sim 0.13$ eV): A  spin-wave theory computation shows that,
at low energies, the 
effect of  the  extended terms in the Hamiltonian is to  renormalize the 
effective $J$  to lower values\cite{sch90}. 

We also computed the staggered magnetic moment in the EHM.
We get a staggered moment roughly 7\% larger than in the HM (see
also Refs.~\cite{hon93,kap92}). This may be
important in view of the disagreement found for this quantity between 
theory and experiment\cite{kap92,kap89}. 

In conclusion we have presented a computation of the IR absorption
spectra due to magnetic excitations in undoped cuprates. We have shown
that these experiments are very sensitive to the underlying magnetic
Hamiltonian. We find that the usual model used to describe the spin
dynamics in cuprates, namely   a HM with nearest-neighbor 
exchange, can not explain the experimental data. 
Instead an EHM 
with further neighbor interactions and with a 4SCE term is in good
agreement with the data. 
To the best of  our knowledge this 
 provides the first quantitative explanation of the puzzling side bands 
in the spectrum. 
In addition, the same model has been found  to  agree  
with other spectroscopic data\cite{lem97ero99}. Further theoretical
work is needed to explore the consequences of 
the 4SCE term in the doped phase. 

 We have been strongly influenced by ideas and views from Eduardo Gagliano 
 who unfortunately passed away during the course of this work. 
 We are in debt with the authors of 
Refs.~\cite{per93,per95,per98,gru96} for providing us with their
original experimental data. 
 Two of us (J. E. and J. L.) were supported by CONICET during part of
this work. J. L. thanks SISSA for  hospitality. 
This work was supported in part by Fundaci\'on Antorchas, 
Fundaci\'on Balseiro,  ANPCYT, INFM PRA (HTSC) and CINECA grant.
Work at Los Alamos Nat. Lab. is sponsored by the US DOE under contract 
W-7405-ENG-36.


\begin{references}
\bibitem{note}
We argue that  the interpretation  of the MR line shape is not conclusive
because: i) The scattering mechanism itself is under
  discussion since resonant effects are important\cite{chu95}. In fact a strong
  dependence of the line shape on the excitation energy has been
  reported\cite{blu96} and one can see that both anomalies change substantially
  from one experimental group to the other. 
 ii) Alternative
  explanations for the width of the two-magnon peak exist without invoking
  the 4SCE term\cite{sin89}.
\bibitem{elsewhere} $\rho_{\rm Raman}$  is for an infinite cluster. 
 This extrapolation  will be discussed elsewhere.
\end{references}

\end{multicols}

\end{document}